\documentclass[final,12pt]{elsarticle}

\usepackage{graphicx}
\usepackage{epstopdf}
\usepackage{amssymb}
\usepackage{hyperref}
\usepackage{lineno}

\journal{Social Networks}
\begin{document}

\begin{frontmatter}



\title{State power and elite autonomy in a networked civil society: \\ The board interlocking of Chinese non-profits}

\author[1,2]{Ji Ma}
\ead{ma47@indiana.edu}
\author[3,4]{Simon DeDeo\corref{cor1}}
\ead{sdedeo@andrew.cmu.edu}
\cortext[cor1]{To whom correspondence should be addressed.}
\address[1]{\footnotesize Indiana University Lilly Family School of Philanthropy, 301 N. University Boulevard, Indianapolis, IN 46202, USA}
\address[2]{\footnotesize Intetix Institute, Beijing 100875, PRC}
\address[3]{\footnotesize Social and Decision Sciences, Carnegie Mellon University, 5000 Forbes Avenue, Pittsburgh, PA 15213, USA}
\address[4]{\footnotesize Santa Fe Institute, 1399 Hyde Park Road, Santa Fe, NM 87501, USA}

\begin{abstract}
In response to failures of central planning, the Chinese government has experimented not only with free-market trade zones, but with allowing non-profit foundations to operate in a decentralized fashion. A network study shows how these foundations have connected together by sharing board members, in a structural parallel to what is seen in corporations in the United States and Europe. This board interlocking leads to the emergence of an elite group with privileged network positions. While the presence of government officials on non-profit boards is widespread, government officials are much less common in a subgroup of foundations that control just over half of all revenue in the network. This subgroup, associated with business elites, not only enjoys higher levels of within-elite links, but even preferentially excludes government officials from the NGOs with higher degree. The emergence of this structurally autonomous sphere is associated with major political and social events in the state-society relationship. Cluster analysis reveals multiple internal components within this sphere that share similar levels of network influence. Rather than a core-periphery structure centered around government officials, the Chinese non-profit world appears to be a multipolar one of distinct elite groups, many of which achieve high levels of independence from direct government control. 
\end{abstract}

\end{frontmatter}

\clearpage

\section{Introduction}

\subsection{Board Interlock and State Power}

When the boards of different organizations have members in common---when their boards \emph{interlock}---they can synchronize both their values and behaviors in the absence of explicit central control~\cite{FENNEMA1978297,mintz1981interlocking,mizruchi1996interlocks,davis1997corporate,dreiling_2011}. Organizations that share key members in this fashion can reap the benefits of network connections and solve coordination problems~\cite{pombo2011outside,faulk2015network}.

Board interlock is widespread in free-market societies, where it emerges in the business sector as means for coordinating decisions and building social influence~\cite{davis1996significance}. In many countries, it spans multiple sectors, and links together the nonprofit, commercial, and political worlds~\cite{ssq_2002,barnes}. In the donation-based charity sector, board interlock helps coordinate of efforts and share of information~\cite{galaskiewicz2006networks}, and enhances both a nonprofit's perceived legitimacy and its capacity to acquire resources~\cite{esparza2013interlocking}. Among ethnic associations, the ``broker function" of board interlock  generates and spreads political trust, helping to build stronger civic communities and strengthening trust towards government~\cite{fennema2001civic}.

Much less is known about the political implications of board interlock under authoritarian governments. For a government concerned with the dangers of independent agents, interlock may be a benefit, because the resulting coordination reduce the independence between organizations and make non-government agents easier to control. However, these benefits exist only if the government maintains control of the most central organizations in the resulting network. If it does not, board interlock may shift from an opportunity to a threat: organizations may not only reap the benefits of coordination, but now do so by coordinating around an independent agent. 

Board interlock is crucial to understanding ``infrastructural'' forms of state power~\cite{mann1984autonomous}. Infrastructural power refers to the capacity of the state to act through civil society, by penetrating, and thereby influencing, its institutions. Infrastructural power is often contrasted with despotic power: the ability of state elites to act without formal negotiations with civil society, through top-down, unilateral action. The coordination enabled by board interlock provides an important means by which a state might amplify infrastructural power---or, conversely, a means by which non-governmental actors may reduce it. 

The world of non-profit foundations in the People's Republic of China provides a key test case for how a central authority confronts the challenges of an emergent network of non-governmental organizations. In short: how does an authoritarian regime deal with the counter-power that may develop when agents of a putative civil society connect together? 

\subsection{How Much Autonomy? A Brief Introduction to Civil Society and the Chinese Nonprofit Sector}

While charities and ``social organizations'' appear early in China's history, the majority were closed down during the Cultural Revolution in the 1960s and 1970s~\cite{ye2003china}. The nonprofit sector only re-emerged during the reform era of the 1980s, as part of the government's push to decentralize and devolve power away from direct state control~\cite{ma2002governance,teets_2013}. In the following decades, the sector has expanded so rapidly that scholars today ask whether or not it represents the rise of a Chinese civil society: a dense network of groups that bring together citizens to accomplish activities outside of government control.

That concept, civil society, has its origin in the 19th Century, when Alexis de Tocqueville connected the early stages of American democracy to the growth of voluntary associations of ordinary citizens for everything from the promotion of temperance to the founding of schools~\cite{tocque}. Ever since, political theorists and sociologists have tried to understand the role that these associations might play in the liberalization of authoritarian regimes and the early stages of democratic rule~\cite{walzer_1992}. The concept of civil society has continued to evolve; in a recent study of the ``illegal'' NGO sector within China, Ref.~\cite{illegal_ngo} quotes Ref.~\cite{foley1996paradox} to describe a neo-Tocquevillian concept of civil society as ``an autonomous sphere of social power within which citizens can pressure authoritarians for change, protect themselves from tyranny, and democratize from below''. For these reasons, hard-line members of the Chinese government are liable to view the very concept of civil society as a ``trap''~\cite{trap}. A central theme of research on the Chinese nonprofit world is thus how autonomous organizations can be in presence of state control~\cite{ma2002governance,alliance_2010}. 

Yet the existence of non-governmental associations does not necessarily imply a civil society in the Tocquevillian mode or even a threat to authoritarian rule. While countries in the West have accepted nonprofits that operate independent of government control, foundations in China must contend with a one-party system potentially intolerant of organizations that might hold it accountable or draw attention to its deficiencies, and that therefore strives to control and monitor it. Concerns about the lack of autonomy in the nonprofit sector have led many observers to talk in terms of \emph{state-corporatism}~\cite{whiting1991politics,ma2002governance}, where the nonprofit sector is an auxiliary and dependent system of the state. In the classic definition of Ref.~\cite{schmitter1974still}, the relevant organizations in state-corporatism parallel those of government agencies, being ``singular, noncompetitive, hierarchically ordered, sectorally compartmentalized, interest associations exercising representational monopolies''. 

In general, civil society can be understood through a paradigm focused either on conflict, or on contingent cooperation. Theories that focus on conflict assume that the state and non-state organizations have goals that are in fundamental tension. These theories leave little room for extensive cooperation between the two sectors. The neo-Tocquevillian conception of civil society is the most explicit form of this conception, while, in the particular case of Chinese non-profits, the idea of civil society as a challenge to state power can be found in Ref.~\cite{kang_graduated_2008} which describes a ``system of graduated control" where the state exerts different control strategies over different types of nonprofits, depending on the level of threat these extra-government organizations are seen to pose.

By contrast, the contingent cooperation paradigm sees non-profits as potential service arms of the state, at times able to implement the state's goals in a more efficient and effective fashion. Spires' 2011 paper~\cite{illegal_ngo} popularized an account of this form, based around the idea of a ``contingent symbiosis" between government and civil society, in which illegal NGOs are allowed to operate as long as they relieve the state's responsibilities for social welfare. Another example is provided by Ref.~\cite{teets_2013}, which describes a ``consultative authoritarianism" that promotes at one and the same time an ``operationally autonomous civil society" and a ``sophisticated authoritarianism that uses more indirect tools of social control".

\subsection{Networked Civil Society}

Because of the power of the Chinese state, research into its nonprofit world tends to focus on how much autonomy can exist in the presence of state control~\cite{ma2002governance,kang_graduated_2008,alliance_2010,hsu2015ngo}. Previous studies have documented the strategies and tactics of individual nonprofits, either through case studies or the identification of qualitative patterns of behavior across multiple cases~\cite{estes1998emerging,saich_2000,yiyi2007autonomy,teets_2013}.

Civil society, however, is more than just the existence, and even the autonomy, of non-governmental organizations. It is how these organizations connect together, in a horizontal fashion, to form something more than a catalog of distinct endeavors~\cite{salmenkari_theoretical_2013}: organized ``multiple, overlapping, and intersecting sociospatial networks of power''~\cite{mann1986sources}.

To understand civil society in China, in other words, we must study not only how the state acts on individual foundations, but also how it interacts with the networks through which these foundations share personnel, information, and resources. The infrastructural power the state exercises may be both enhanced, and dissipated, by the horizontal connections between the organizations it penetrates. Board interlock is one of the primary mechanisms for this self-organization to take place, and yet we know next to nothing about how this process has unfolded, and the implications of this evolution for civil society in twenty-first century China.

We will study the Chinese state-society relationship by looking at the evolution of the non-profit board interlock network. To do this, we draw on a large dataset of officially-registered nonprofit foundations. This dataset records not only important information about each foundation, but also the list of board members, enabling us to construct the board interlock network. Our analysis can then operate at two levels simultaneously: (1) at the level of the individual foundation, and (2) at the level of the network, where edges between foundations are defined by the sharing of board members.

At the level of individual foundations, our data show the high level of presence of government officials on foundation boards. Examination of how the number of government officials varies by working areas and foundation type shows how the presence of government officials correlates with activities, and legal status, that the government is expected to be most concerned to control.

At the network level, we find that board appointments connect together a significant fraction of legal Chinese foundations into a single network. Our results show the existence of network elites, associated with business entrepreneurs and their foundations, that form preferential ties to each other. This subnetwork appears to preferentially exclude government officials from its most central nodes, providing evidence that the network acts in part to frustrate the magnification of state control that might be expected to arise in the presence of board interlock.

The sharing of board members not only connects foundations together, it also appears to preferentially connect them into clusters: groups of nodes that, taken together, tend to preferentially connect to each other rather than the other nodes in the network. This phenomenon has been studied quantitatively by Ref.~\cite{heemskerk2016corporate}, for the case of the board interlock network among large corporations. In that study, the authors were particularly interested to determine whether the interlock network showed either a classical ``core-periphery'' structure based around a single hegemon, or evidence for a the existence of a more ``multipolar global order''.

Such an analysis has a natural analog to the question of the extent of civil society within a nation: do foundations interlock with a government-controlled core, or do they associate into independent structures with central positions that rival that of any putative hegemonic core? We find strong evidence for the latter, detecting distinct clusters at the very center of the network. These clusters show a strong bimodaility in levels of government control, either preferentially excluding, or including, government officials. The Chinese non-profit world is a multipolar one, with different groups at the center showing distinct relationships to government control.

Taken together, our findings suggest the emergence of a form of network autonomy that exists despite high levels of individual-level government appointments to nonprofit boards. At the same time, the association of this network autonomy with the business elite---rather than ``ordinary'' citizens---means that this autonomy may not lead to the kind of pluralism associated with a Tocquevillian civil society.

\section{Methods}

\subsection{Dataset and Network Construction}

Our primary dataset is the Research Infrastructure of China Foundations (RICF~\cite{ma_database_2015}; the underlying data is available at Ref.~\cite{sci-data}, and additional, processed files including the networks derived for this analysis are available at~\cite{simon}). The RICF database contains the records of the 3,344 legally-registered foundations within mainland China between 1981 and 2013. Information about each foundation is drawn from six different sources, including both official government reports and information submitted to the government, or reported on websites, by the foundations themselves. Comparing RICF's counts to other reference sources, the RICF's data appears to be at least 90\% complete. The data is restricted to foundations allowed to conduct fundraising; while these foundations are outnumbered by the much larger number of less formal ``associations'', they are the most developed form of non-profit institution, and control just over 76\% of all non-profit funding in the nation.\footnote{The 3,344 foundations in the RICF database control approximately 35.3 billion RMB (B$\yen$); this amounts to a large fraction of the 45.9 B$\yen$ of non-profit funding, tracked by the Ministry of Civil Affairs of China in 2013, for all ``associations'' and related classifications~\cite{bulletin}.}

RICF strives to be as comprehensive as possible; it includes foundations that may, for example, may be essentially defunct. Within RICF data is a subset of foundations that have undergone an evaluation process, which rates the foundations according to a set of criteria including governance structure, financial transparency, and program effectiveness. Foundations which receive a ``3A'' or above (3A+) are considered to have passed this evaluation. Because the 3A+ evaluation includes checks on transparency and reporting, we expect the data associated with 3A+ foundations to be more reliable; because it also includes checks on effectiveness and governance, we expect these foundations to be more active. Comparing our analyses with the 3A+ set allows us to test for unexpected sensitivities to both data quality, and foundation activity levels, in the database as a whole.

For each foundation, RICF logs the names, gender, and date of birth of the board members. This allows us to resolve name collision and therefore construct the board interlock network: two foundations are connected when they share one (or more) board members. For simplicity in this analysis, our network is unweighted: we consider only the presence or absence of at least one shared member, and do not distinguish whether links are created by sharing presidents, secretaries, or ordinary members. Extensions to the study of weighted ties are certainly possible: one could, for example, consider weighting the edge by the number of board members shared. Such a weighting would add additional methodological complexities, however, since foundations with larger boards will have the ability to form stronger ties with each other. This may or may not correctly represent the underlying social dynamics: if two foundations with very large boards share two members, it may not make sense to represent them has having stronger ties to each other compared to two foundations with very small boards and only a single member in common. One could attempt to normalize the weights by board size; in the case, however, that the two foundations so connected have different sizes, we now have asymmetric edge weights. More sophisticated models yet could be constructed, on the basis of a probabilistic model for connection formation: for example, one could consider edge weight relative to a null model where ties are formed at random. In this analysis, for both simplicity and to allow direct comparison to prior literature in other areas of the world, we follow the standard choice of unweighted edges.

The RICF data also contains the date of incorporation for each foundation, allowing us to study how the final network assembled over time. Inconsistencies in historical reporting rates and availability of data make it difficult, if not impossible, to produce an exact history of entrances and departures over time. We can, however, approximate this process by seeing how the network would have assembled if the foundation's board had remained unchanged to the present. 

This is only an imperfect tracer of the more detailed question that includes both the formation of a foundation, and the ways in which the network might be altered by the addition or removal of board members. However, prior work suggests that this approximation may not be that bad. A study of non-profits in Spain found that initial boards were usually assembled by the founders, and that while nonprofits needed a period of development to attract outsiders as potential board members~\cite{de2009determinants}, this early growth appeared to stop when the non-profit reached maturity. A study in the United Kingdom found that the majority of nonprofit boards remained unchanged on timescales of three years, and that it was difficult for to recruit new board members~\cite{cornforth2002change}. Little is known about the Chinese case, however; where relevant, we draw attention to cases where ``the structure at time $t$, given board compositions in 2015'' is only an approximation, at best, to ``the structure at time $t$, given board compositions at time $t$''. In general, our dynamical analysis is limited to asking questions of the form ``when did organizations responsible for the current trend join the system?''. 

\subsection{Variables}

Four critical variables help us characterize foundations at the individual level.

\textit{Public Fundraising} vs.\ \textit{Non-Public Fundraising}. The main legal distinctions in the Chinese nonprofit sector governs the scope of fundraising. ``Public fundraising'' foundations are allowed to raise money from the general public---for example, through fund drives and advertising---while ``non-public fundraising'' foundations may not (as shorthand, we refer to these as ``public'' and ``non-public'' hereafter). Moreover, public foundations are further constrained by geography; ``central-level'' foundations may raise funds at the national scale, while province-level and city-level foundations are restricted geographically. The Chinese Charity Law, effective from 1 September 2016, will nullify the distinction between public and non-public fundraising; however, the ability of organizations to raise funds from general public will still be controlled by license issued from the state.\footnote{Xinhua News Agency: NPC hopes charity law can help poverty fight, available at \url{http://news.xinhuanet.com/english/2016-03/09/c_135172544.htm}, last accessed June 14, 2016.}

\textit{Politically Sensitive} vs.\ \textit{Politically Non-Sensitive}. We supplement the RICF by hand-coding the foundations' mission statements by whether or not they are involved in a potentially controversial or politically sensitive area (``sensitive'' vs.\ ``non-sensitive''). Tracking this variable allows us to look for systematic attempts to selectively control certain topics. If a foundation has one or more of the following characteristics it is coded as ``sensitive," otherwise it is coded as ``non-sensitive" \cite{illegal_ngo, kang_graduated_2008, dai2008policing, jiao2001police}:

\begin{enumerate}
	\item Involving advocacy, \emph{e.g.}, human rights, labor issues, and environmental policy.
	\item Involving international affairs, \emph{e.g.}, programs promoting international cultural exchanges.
	\item Involving religious or ethnic issues, \emph{e.g.}, Christian activities and Tibet issues.
	\item Involving the police or the legal system, or related ``social stability'' (\textit{weiwen}) issues.
\end{enumerate}

In order to control the researchers' bias toward coding, two assistants, who were unaware of the research purpose, were asked to independently classify foundations, solely according to the information provided in their mission statements. Discrepancies between the two assistants were finalized by a third person who is a doctoral candidate in China studies.

\textit{State Power}. Our main tracker of state penetration into the non-profit sector is the number of government officials in senior management positions. The presence of government officials on a foundation board is a clear mechanism by which the state can exercise control; at the same time the state, at least explicitly, forbids government officials from serving on these boards. Official law (Article 23) is that ``principals'' (the board chair (president), deputy chair, or secretary general) should not be currently employed by the state.\footnote{``Regulations on the Management of Foundations'' (4 February 2004), Article 23; original text available at \url{http://www.mca.gov.cn/article/yw/shjzgl/fgwj/201507/20150700850200.shtml}, see English translation at \url{http://www.cecc.gov/resources/legal-provisions/regulations-on-the-management-of-foundations-chinese-text}, last accessed 18 April 2016.}

Law against direct government involvement are commonly violated. The standard annual reporting forms even asks foundations to report the number of principals who are government staff; a non-negligible fraction (18\%) report non-zero numbers of current government officials. As a different measure of state influence, we hand-coded the 3A+ foundations, noting whether or not the board president is a current or retired government official. Hand-coding is a difficult and laborious task; because it can be difficult to ascertain the current status of individual, our 3A hand-coding includes both retired and currently-serving government officials, and is thus not directly comparable to the self-reporting set.

We thus count the number of government officials in three different ways. The RICF counts the number of \textit{current} government officials who serve as principals; it also counts the number of simultaneously \textit{retired} and \emph{senior} government officials who serve as principals; and (for a hand-coded subset) it counts the number of either current or retired government officials who serve as the board president. Note that the Article 23 law does not forbid retired government officials from serving as board principals, and these people may well still represent government influence~\cite{li98}.

\textit{Registration Level}. Depending on their scope of operation, foundations may be registered at the city level, the province level, or the central level. Central-level registrations enable the foundation to operate on a national scale. Registration level gives us information on both the scope of foundation operation, and also allow us to look for preferential interlocking as a function of both scope (\emph{e.g.}, do central-level foundations preferentially connect to each other) geography (\emph{e.g.}, do same-city registrations connect preferentially versus different-city).

\section{Results}

\subsection{Individual-Level Statistics and the Presence of State Power}

\begin{table}
\begin{tabular}{c|c}
Total Numbers & Yearly Income \\ \hline 
\\
\begin{tabular}{l|l|l}
& Non-sensitive & Sensitive \\ \hline
Public & 954 (28\%) & 356 (11\%)  \\
Non-Public & 1913 (57\%) & 121 (3\%) \\
\end{tabular} &
\begin{tabular}{l|l|l}
& Non-sensitive & Sensitive \\ \hline
Public & 13.9 B$\yen$ & 3.2 B$\yen$  \\
Non-Public & 17.5 B$\yen$ & 0.7 B$\yen$ \\
\end{tabular}
\end{tabular}
\caption{Foundation status, activity, and yearly budgets (in billions of \emph{yuan}; $1\ \mathrm{B}\yen\approx150\ \mathrm{M}\$$ in 2016) for the 3,344 foundations in our database. The majority of the foundations are concerned with neutral (non-sensitive) activities, and the majority are restricted to private fundraising. Despite the fact that non-public foundations can not raise money from the general public, they actually control the majority of nonprofit revenue in the network.\label{counts}}
\end{table}Table~\ref{counts} shows the breakdown of the foundations in the RICF data. Collectively, the 3,344 foundations in our data have a combined income of 35.3 billion RMB; approximately 5.5 billion USD at current exchange rates. Paralleling Tocqueville's accounts of the diverse initiatives of citizens in 19th Century America, foundations in contemporary China range in activity in everything from the promotion of the board game \emph{w{\' e}iq{\' i}} (Go) to legal aid for the indigent. The majority are restricted to non-public fundraising, and work in non-sensitive areas. These non-public foundations control just over half of the total yearly nonprofit revenue in the country. 

\begin{table}
\begin{tabular}{c|c}
Official or Retired-Official President & Official Principal  \\
(Hand-coded subset) & (Self-reports) \\ \hline \\
\begin{tabular}{l|l|l}
& Non-Sensitive & Sensitive \\ \hline
Public & $74\%\pm3\%$ & $79\%\pm5\%$ \\
Non-Public & $41\%\pm3\%$ & $29\%\pm9\%$ \\
\end{tabular} &
\begin{tabular}{l|l|l}
 & Non-Sensitive & Sensitive \\ \hline
Public & $34\%\pm1\%$ & $54\%\pm3\%$ \\
Non-Public & $9\%\pm2\%$ & $19\%\pm4\%$ \\
\end{tabular}
\end{tabular}
\caption{Government presence on foundation boards is widespread, but variable. Despite explicit laws against the practice, foundations often have government officials as board principals. Hand-coding of a subset of 520 foundation presidents (left panel) shows that the practice of incorporating current or retired government officials is widespread. In both hand-coded data, and self-reports in annual filings (right panel), foundations able to raise funds from the general public, and foundations concerned with government unfavorable activities, are more likely to be controlled in this fashion. \label{control}}
\end{table}
Despite laws to the contrary, our results confirm persistent and high levels of state involvement in the governance of foundations. Table~\ref{control} shows the relationship between fundraising scope and activity, relying on both the hand-coded 3A subsample that tracks the affiliations of board presidents, and official self-reports in the full database. Because the 3A hand-coding includes retired officials, the two methods track slightly different phenomena.

Both methods confirm that the presence of government officials is less for non-public foundations. The government is most involved in foundations that are able to raise funds from the general public. (Self-reports, but not our hand-coding, show additional supervision when the foundation itself is associated with sensitive activities. Because this signal does not appear in the hand-coded subset, it may be best-explained by differing incentives: public foundations involved with sensitive activities may be more willing to report government officials on their boards, despite the fact that this violates Article 23.)

\subsection{Board Interlocking: Super-Connectors and Elites}

\begin{table}
\begin{tabular}{l|l|l}
& Public & Non-Public \\ \hline
Central & $82\%\pm1\%$ & $72\%\pm1\%$ \\
Provincial & $47\%\pm2\%$ &$36\%\pm2\%$ \\
City & $67\%\pm1\%$ &$24\%\pm1\%$ \\
\end{tabular}
\caption{The fraction of nodes of each type that share board members with other foundations. The board interlock network extends to a significant fraction of all Chinese non-profits. Both public and non-public fundraising foundations are highly integrated into the overall network, with central-level foundations able to raise from the general public the most connected of all. \label{integration}}
\end{table}
\begin{figure}
\begin{center}
\includegraphics[width=6in]{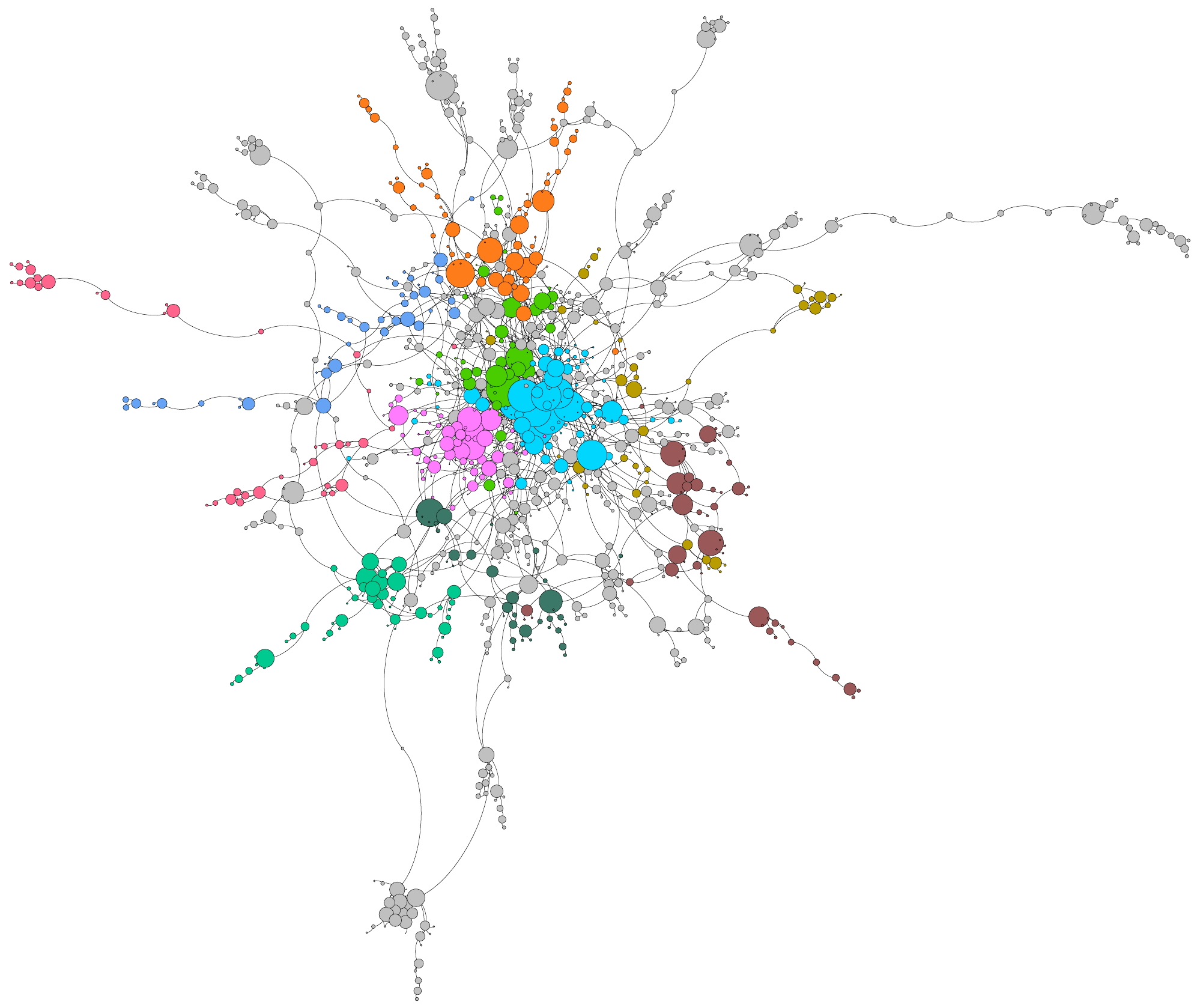}
\end{center}
\caption{The giant component of the board interlock network, containing 1022 nodes and 1626 edges; 75\% of the nodes with non-zero degree, and 30\% of the full database. A simple spring loaded network layout algorithm allows us to visualize which nodes are tightly coupled to many others (end up at the center), and which are connected to the main network by only a small number of links to peripheral nodes.  At the center are a small number of interlinking elite super-connectors with high degree (Fig.~\ref{degree}; Fig.~\ref{rc}). Node size is scaled by PageRank; node colors label the top ten largest clusters found using the Louvain algorithm (see Table~\ref{clusters}).\label{bi}}
\end{figure}Board interlock is widespread. A significant fraction of the foundations are integrated into the network: of the 3,344 foundations, 1,411 (42\%) share board members with at least one other foundation, for a total of 1,863 links. As shown in Table~\ref{integration}, foundations at the central level are the most likely to be connected. Both public and private foundations show significant network integration. A large fraction of this network is connects together, into a single giant component that contains 1,022 foundations (see Fig.~\ref{bi}). 

While the existence of board interlock parallels the dominant corporate cases studied in the West, there are significant differences. Most notably, the network is not small world: the average path length between nodes in the giant component is $7.71$, and the network diameter is $27$; in both cases much larger than the corporate board interlock seen in the United States~\cite{davis2003small}. 

The board interlock network in Chinese foundations, in other words, has a tendency to isolate nodes from each other. Even if we restrict to the 77 central-level foundations in the giant component, the diameter remains large (9 steps); by comparison, the network of Fortune 800 firms in the 1970s, nearly ten times larger, had a diameter of five~\cite{levine1977network}. Chinese foundations connect to each other, but the existence of these long paths shows that board interlocks are fundamentally limited in their ability to coordinate action on the very largest scales.

\begin{figure}
\begin{center}
\includegraphics[height=2in]{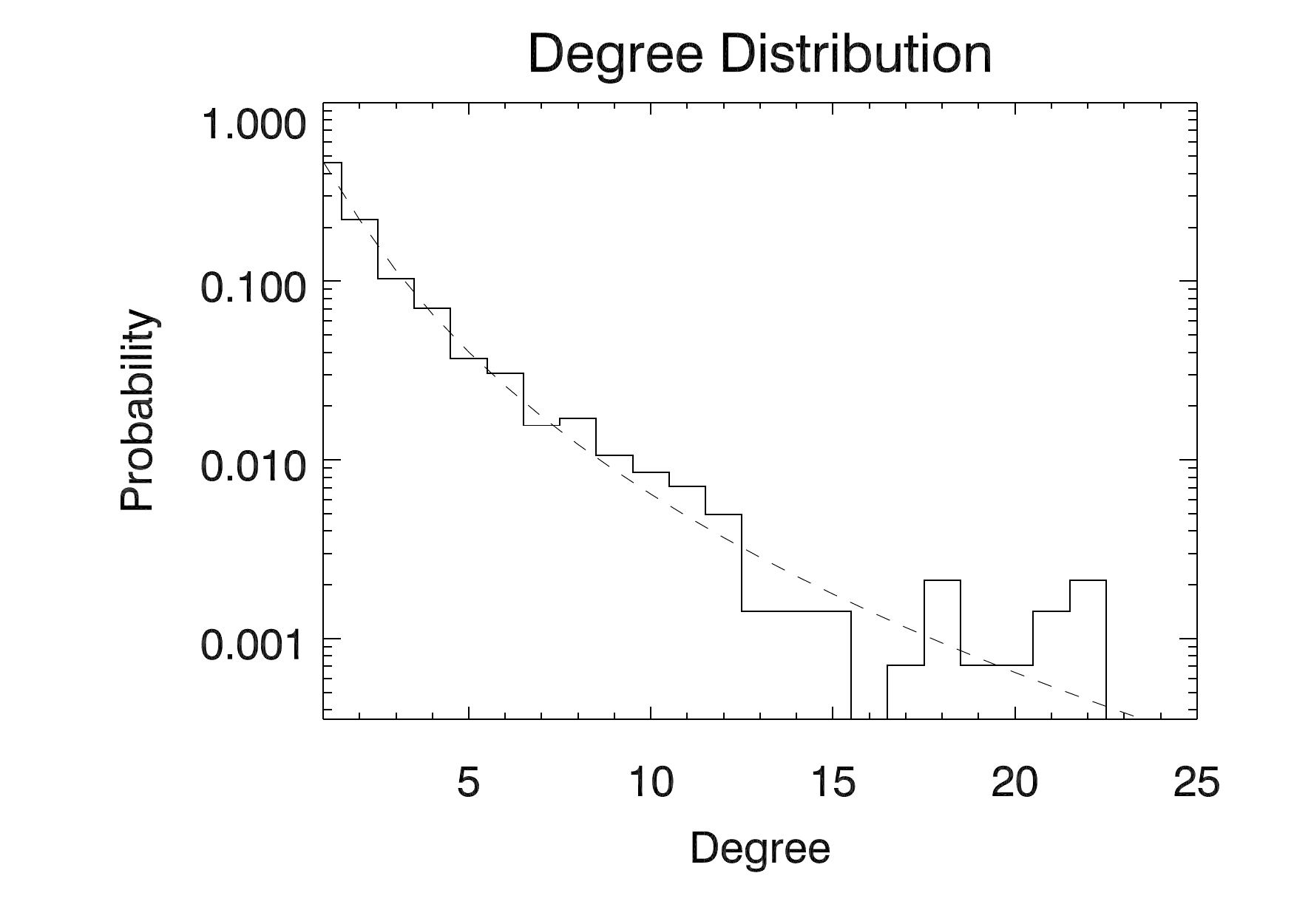}
\end{center}
\caption{The degree distribution of the board interlock network. The distribution is log-normal (dashed-line fit), and a small fraction of the nodes have unusually high degree.\label{degree}}
\end{figure}
While the network has few shortcuts and hubs that connect otherwise distant nodes, it is also the case that a small number of foundations have very high degree---they share an unusually large number of board members with other foundations. We show the network degree distribution in Fig.~\ref{degree}. The existence of these ``super-connectors'' can be empirically confirmed by testing for heavy-tailed degree distributions; standard methods strongly prefer a log-normal distribution to both an exponential (\emph{i.e.}, random-graph) and power-law fit~\cite{clauset2009power,powerlawfit}.

\begin{figure}
\begin{center}
\includegraphics[height=2in]{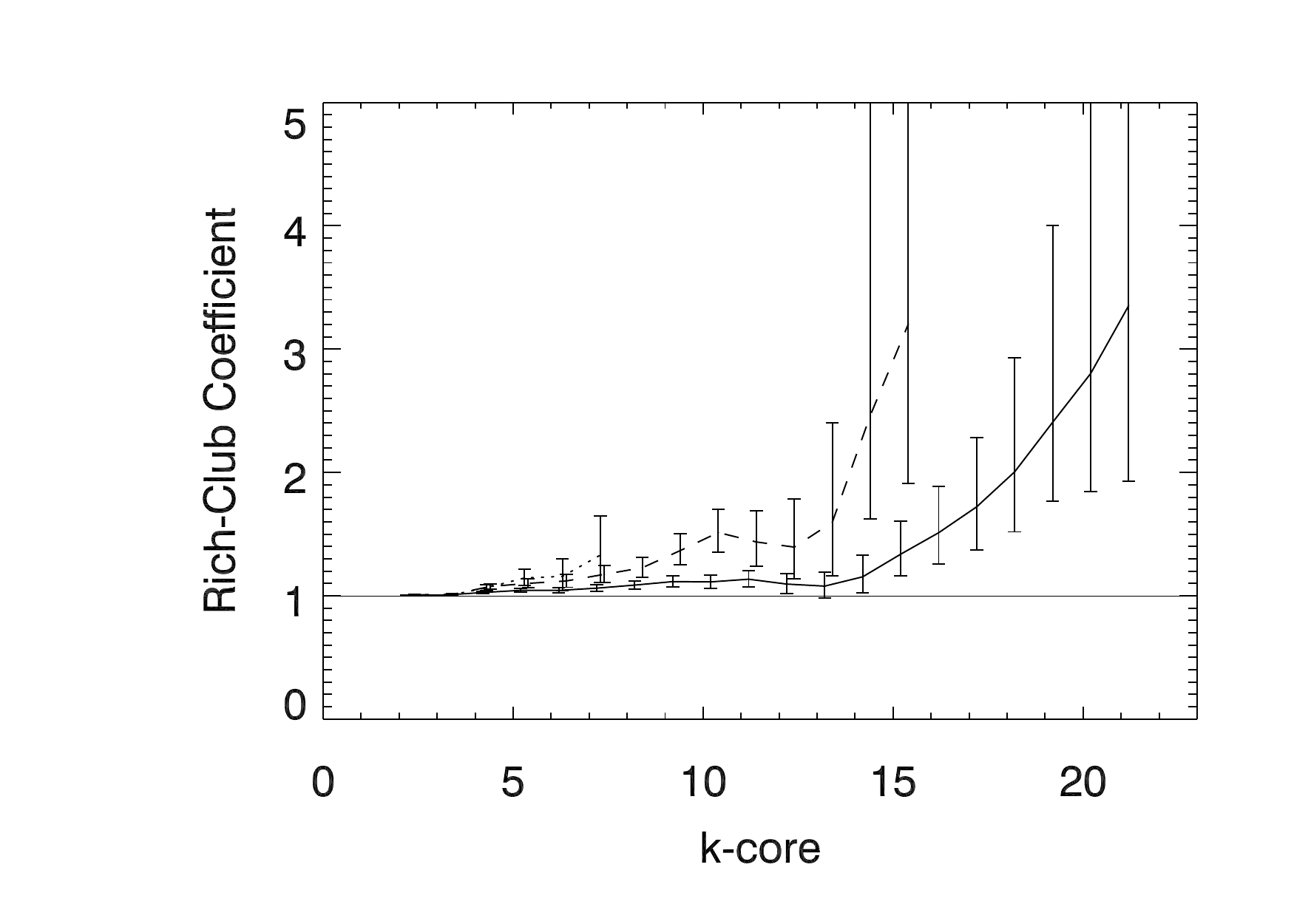}
\end{center}
\caption{The rich-club coefficient as a function of $k$-core. High-degree organizations (``network elites'') preferentially connect to other network elites, particularly in the non-public network. This is apparent in how the rich-club coefficient grows as a function of $k$-core level for the full network (solid line), the public network (dotted line), and the non-public network (dashed line); 95\% confidence ranges are shown. Nodes with high degree are far more likely to connect to each other, compared to a null model that preserves the degree distribution but otherwise breaks interlock correlations~\cite{mcauley2007rich}. Taken separately, the non-public network shows a stronger rich club effect than the network as a whole. Meanwhile, the majority of high-degree links in the public network are due to cross-links with the non-public foundations; the maximal degree for the public network in isolation is much smaller. \label{rc}}
\end{figure}
These super-connectors appear to preferentially connect to each other, suggesting the existence of highly connected elite groups (the ``rich-club phenomenon", first noted by Ref.~\cite{zhou2004rich}). Fig.~\ref{rc} shows the rich-club coefficient in the full, the public, and the non-public subnetworks. To read this figure, first follow the solid line; that this line is rising as a function of $k$ shows that, in the full network, nodes with higher degree are more likely to connect to each other than in a null model that preserves the node degree distribution but otherwise shuffled connections. Now follow the dashed line, which considers just the internal connections of the non-public subnetwork. This line shows that this same phenomenon, seen in the full network, is even stronger here: non-public elites are even more likely to preferentially connect within this subnetwork. Finally, note the (very short) dotted line. This shows at best weak evidence for a rich-club effect when restricting to the public network. The line is much shorter, because the public network, considered in isolation, has few high-degree nodes: if a public foundation has high-degree in the full network, it is usually because it connects to non-public foundations.

If we code nodes by the board president's profession, we find that the highest degree nodes include both the government and the business elite. Of the ten highest-degree foundations, six are associated with businesses, four with government. For example, the most connected foundation is the YouChange China Social Entrepreneurs Foundation, set up to encourage philanthropic giving by wealthy entrepreneurs; the second is the Forbidden City Cultural Heritage Conservation Foundation, which manages the state-owned historical treasure. The top ten most connected board members are also a mixture of both government and business elites; three are business elites, six are current government officials, including members of the National People's Congress, the CPPCC, and the Guangzhou People's Congress; and one is a retired government official.

Considering the public and non-public networks separately allows us to see how different groups dominate. In the public network, the highest-degree nodes are nearly all associated with the government: of the top ten highest degree foundations, only two have a president with a non-governmental background (a television celebrity, and the head of a hospital). By contrast, six of the ten highest degree nodes in the non-public network are associated with business elites. The highest degree node, for example, the YouChange Foundation, is run by the businesswoman Ping Wang, whose background is in international finance and law.

\begin{table}
\begin{tabular}{l|l|l|l}
& Central & Province & City \\ \hline
Central & $\times\textbf{2.1}\pm\textbf{0.1}$ & $\times0.73\pm0.03$  & $\times1.2\pm0.2$ \\
Province (same) &  --- & $\times\textbf{7.05}\pm\textbf{0.08}$ & $\times4.7\pm0.5$ \\
Province (different) & --- & $\times0.42\pm0.01$ & $\times0.46\pm0.05$ \\
City (same) & --- & --- & $\times\textbf{14.6}\pm\textbf{0.9}$  \\
City (different) & --- & --- & $0$ \\
\end{tabular}
\caption{The number of links found between nodes of each type, compared to a degree-preserving null. Foundations in the same region, and at the same registration level, tend to cross link. Centrally-registered foundations are more than twice as likely to connect to each other than in the null; province-level foundations connect to others in the same province at rates seven times higher than the null, and city-level foundations are the most cross-linked of all, linking to other city-level foundations in the same place at rates 14 times higher than expected.\label{geography}}
\end{table}
Board interlock is influenced by both geography and registration level; see Table~\ref{geography}, paralleling classic results for corporate networks, co-located foundations are far more likely to connect~\cite{stuart1998network,sorenson2001syndication,owen2004knowledge}. We also find evidence for preferential connections between sensitive foundations; there are $70\%\pm10\%$ more links between sensitive foundations than found in the null. However, public- and non-public foundations appear to intermix freely and preferences (though detectable) are weak; public foundations are only $16\%\pm3\%$ more likely to link to each other than null, and only $17\%\pm3\%$ less likely to link to non-public foundations.


\subsection{Network Effects: Penetration of State Power}

\begin{table}
\begin{tabular}{l|l|l|l}
& Public & Non-Public & Joint \\
Predictor & Network & Network & Network \\ \hline
(Legal Status) & & & \\
Central Level & $\mathbf{0.72\pm0.15^{\star\star\star}}$ & $\mathbf{0.90\pm0.16^{\star\star\star}}$ & $\mathbf{1.6\pm0.2^{\star\star\star}}$ \\
Evaluation 3A+ & $\mathbf{0.33\pm0.12^{\star\star\star}}$ & $\mathbf{0.55\pm0.12^{\star\star\star}}$ & $\mathbf{0.66\pm0.12^{\star\star\star}}$ \\ 
Public & --- & --- & $-0.1\pm0.1$ \\ \hline
(State Power) & & & \\
Current Official & $\mathbf{0.17\pm0.08^{\star}}$ & $\mathbf{-0.44\pm0.13^{\star\star\star}}$ & $-0.15\pm0.11$ \\
Retired, Senior Official & $0.0\pm0.1$ & $0.1\pm0.2$ & $0.2\pm0.2$ \\ \hline
(Intrinsic) & & & \\
Board Size (z) & $\mathbf{0.31\pm0.04^{\star\star\star}}$ & $\mathbf{0.28\pm0.04^{\star\star\star}}$ & $\mathbf{0.52\pm0.04^{\star\star\star}}$ \\ 
Income (z) & $\mathbf{0.19\pm0.04^{\star\star\star}}$ & $\mathbf{0.16\pm0.04^{\star\star\star}}$ & $\mathbf{0.40\pm0.04^{\star\star\star}}$ \\ 
Age (z) & $-0.03\pm0.05$ & $0.00\pm0.04$ & $-0.06\pm0.04$ \\
Sensitive Area & $-0.10\pm0.08$ & $\mathbf{0.33\pm0.16^{\star}}$ & $0.13\pm0.11$ \\ \hline
& $R=0.42$ & $R=0.34$ & $R=0.44$
\end{tabular}
\caption{Predictors of node degree, in public and non-public fundraising networks, and in the joint network, in a multiple linear regression. Non-public foundations preferentially exclude current government officials from the highest degree nodes, even when controlling for other variables. (z) indicates z-scored transformed real variables; all other variables are binary for presence/absence. $\star$ superscripts label significance: $\star$ ($p<0.05$); $\star\star$ ($p<0.01$); $\star\star\star$ ($p<0.001$). \label{all}}
\end{table}
We next consider how the presence or absence of government officials predicts node degree. We use a multiple linear regression model, with node degree as the dependent variable and nine independent variables: three variables describing the node's legal status (public vs. non-public; registration level; evaluation level), four ``nuisance'' variables describing the intrinsic properties of the foundation (board size, income, sensitive area, and age), and two variables operationalizing the state power: (1) current government officials, and (2) retired, senior-level government officials, the two fields in the main RICF database. We consider both the full network, and the two public and non-public networks separately. The results are shown in Table~\ref{all}. 

The most surprising results concern the relationship between the presence or absence of government officials, and node degree. Foundation degree is (weakly) positively correlated with government presence in the public foundations: nodes with higher degree are more likely to have government officials. However, in the non-public network, high degree is strongly (and significantly, $p<10^{-3}$) correlated with reduced government presence. When the two networks are joined together, the two effects compete against each other, partially canceling out. Significantly, we find no correlation for the presence or absence of retired senior officials; only the absence of currently-serving officials is predicted by node degree. To allow comparisons between different networks, we use $z$-score transformed data; for example, in Table~\ref{all}, the presence of a current official as board principal in the non-public network shifts node degree down by roughly $0.44$ standard deviations (controlling for other factors), where a standard deviation is measured for all the nodes in the non-public degree distribution.

\begin{table}
\begin{tabular}{l|l|l|l}
& Public & Non-Public & Joint \\
Predictor & Network & Network & Network  \\ \hline
(Legal Status) & & & \\
Central Level & $\mathbf{1.06\pm0.26^{\star\star\star}}$ & $\mathbf{1.0\pm0.5^{\star}}$ & $\mathbf{2.6\pm0.4^{\star\star\star}}$ \\ 
Public & --- & --- & $-0.3\pm0.3$ \\ \hline
(Government Control) & & & \\
President Official & $0.06\pm0.25$ & $\mathbf{-0.68\pm0.37^{\star}}$ & $\mathbf{-0.65\pm0.34^\star}$ \\ \hline
(Intrinsic) & & & \\
Board Size (z) & $\mathbf{0.42\pm0.09^{\star\star\star}}$ & $\mathbf{0.60\pm0.19^{\star\star}}$ & $\mathbf{0.81\pm0.16^{\star\star\star}}$ \\ 
Income (z) & $\mathbf{0.28\pm0.09^{\star}}$ & $\mathbf{0.38\pm0.18^{\star}}$ & $\mathbf{0.75\pm0.15^{\star\star\star}}$ \\ 
Age (z) & $-0.04\pm0.09$ & $0.0\pm0.2$ & $0.0\pm0.15$ \\
Sensitive Area & $-0.40\pm0.20$ & $0.4\pm0.6$ & $-0.4\pm0.4$ \\ \hline
& $R=0.48$ & $R=0.32$ & $R=0.44$
\end{tabular}
\caption{Predictors of node degree, in public and non-public fundraising networks, and in the joint network; hand-coded 3A.  Use of a different, hand-coded dataset for government presence confirms the results of Table~\ref{all}: non-public foundations preferentially exclude current government officials from the highest degree nodes. The ``president official'' code includes retired officials.\label{3a}}
\end{table}
Because Table~\ref{all}'s results rely on self-reports, it is possible that these effects may be driven in part due to differences in self-reporting. We can check for this effect by using the hand-coded 3A subset; these results are shown in Table~\ref{3a}; because the 3A hand-coding tracks only presidents, and includes both current and retired officials, our results here are not strictly comparable to the Table~\ref{all} case. The smaller size of this set also means that our signal-to-noise is lower. In this subset, the weaker correlation in the public data is no longer detectable, but we still see the negative correlation between network position and state presence.

\subsection{Network Clusters: Multipolarity}

Following Ref.~\cite{heemskerk2016corporate}, we use the Louvain algorithm~\cite{blondel08} to detect clustering among the foundations in the network. The Louvain algorithm groups nodes into clusters that maximize the total modularity, $Q$; informally, the modularity for a cluster is defined the fraction of edges within that cluster, minus the fraction of edges expected under a degree-preserving null model. When the modularity is close to one, nodes in a cluster are strongly connected to each other, and only weakly connected to nodes outside. When the modularity is close to zero, by contrast, there is little evidence for the existence of clusters of nodes that preferentially interconnect.

The foundation network shows strong modularity: the average $Q$ for the network is 0.816, and we find a total of 29 clusters. The top ten clusters by membership include just over 50\% of all foundations in the network. They are shown, using different colors, in Fig.~\ref{bi}. The strong modularity of the network points to the existence of structurally independent groups, a precondition for multipolarity. However, it is possible for clusters to exist in the presence of a strong core-periphery structure: a core hegemonic cluster could be at the center of a set of more peripheral clusters. To test for this effect, we consider the average degree of nodes within the cluster, as well as the total PageRank of each cluster; PageRank provides an independent measure of network centrality, based on a random-walker model, and can be interpreted as a consensus measure of attention~\cite{bradi} or power~\cite{brush}.

The results argue for the multipolarity hypothesis. While the largest cluster has higher average degree, and larger PageRank than the others in the top ten, the effect is not strong, and power is shared remarkably evenly among the largest clusters. We can quantify this using the Gini coefficient, which for the top-ten clusters is 19\%, where zero is perfect equality. Put another way, the distribution of PageRank among the top ten clusters is more equal than the famously egalitarian income distribution in Denmark (Gini coefficient 25\%~\cite{cia}), and far less than seen in PageRank distributions found in networks among pages in the world-wide web~\cite{bradi}. The existence of multiple clusters with high degree and high PageRank is consistent with the rich club effect found above: these highly central clusters preferentially link within themselves.

\begin{table}
\begin{tabular}{l|l|l|l|l}
Cluster & Size (Nodes) & Average Degree & PageRank & Government Presence \\ \hline
1 & 91 & $5.4\pm0.6$ & 11\% & $\mathbf{2.1\%^{\star\star\star}}$ {\bf (low)} \\
2 & 68 & $3.3\pm0.4$ & 7.0\% & $\mathbf{2.9\%^{\star\star\star}}$ {\bf (low)}  \\
3 & 66 & $3.9\pm0.4$ & 6.5\% & $11\%$ \\
4 & 55 & $4.3\pm0.6$ & 5.7\% & $15\%$ \\
5 & 53 & $2.7\pm0.4$ & 5.2\% & $\mathbf{30\%^{\star\star\star}}$ {\bf (high)} \\
6 & 50 &  $3.1\pm0.4$ & 5.1\% & $16\%$ \\
7 & 46 & $2.6\pm0.2$ & 4.2\% & $24\%$ \\
8 & 43 & $2.9\pm0.3$ & 4.0\% & $\mathbf{28\%^{\star}}$ {\bf (high)} \\
9 & 39 & $2.7\pm0.2$ & 3.7\% & $\mathbf{28\%^{\star}}$ {\bf (high)} \\
10 & 38 & $2.6\pm0.4$ & 3.5\% & $\mathbf{34\%^{\star\star\star}}$ {\bf (high)} \\
\end{tabular}
\caption{Size (in nodes), average degree, total PageRank, and rates of government presence on boards, in the ten largest clusters of the foundation network. Network power, as measured by PageRank, is shared reasonably equally among the clusters. Six of the ten networks show anomalously high or low levels of government presence compared to the baseline rate of 17\%. \label{clusters}}
\end{table}
Finally, for each cluster, we compute the fraction of nodes within the system that have government officials on the board. This allows us to test whether or not the clusters are defined by characteristic levels of government involvement. For the top ten clusters, we find strong evidence for bimodality in levels of government control. Clusters are much less likely to show ``average'' levels of government presence on boards, and tend to either extreme, either strongly excluding government officials, or having far more than expected given the base-rate. Cluster-by-cluster in the top ten, we have four cases where rates of government presence are either anomalously high, or low, at the $p<10^{-3}$ level, and two more at the $p<0.05$ level. The largest and most central cluster of all has the lowest rates of government presence: 2.1\%, eight times lower than the expected (null) rate of 17\%. The existence of these extreme values in either direction allow us to reject the null hypothesis---that rates of government presence are independent of cluster membership---at $p\ll10^{-6}$ in the standard Fisher test. A Bayesian model, which explicitly models clusters as draws from one of two binomials with different values for the probability of government control, is strongly preferred over a single binomial distribution at similar levels. These relationships are shown in Table~\ref{clusters}.

Taken together, these results suggest that the foundation network is a fundamentally multipolar structure, where clusters have roughly equal levels of network power and are subject to distinct levels of government control. These clusters are further defined both by geography (since foundations with provincial or city-level registrations tend to strongly associate, see Table~\ref{geography}) and public/non-public status.

\subsection{Longitudinal Analysis: the Emergence of Autonomy}

How did these network effects evolve over time? Because board membership changes over time, a full answer to this question would require knowledge of the dates of both joining and departure for each member. In the absence of this information, we can conduct an analysis of the evolution of this network using the board compositions observed in 2015. This amounts to a longitudinal analysis of civil society growth rather than a direct study of network dynamics since we expect, particularly for the older organizations, some changes in board composition due (at the very least) to retirement. While this is, on short timescales and under the assumption that most organizations have relatively stable compositions year to year, a potentially good approximation, such an analysis is better understood as answering the question ``when did the organizations responsible for the current structure join the network''.

\begin{figure}
\begin{center}
\includegraphics[height=2in]{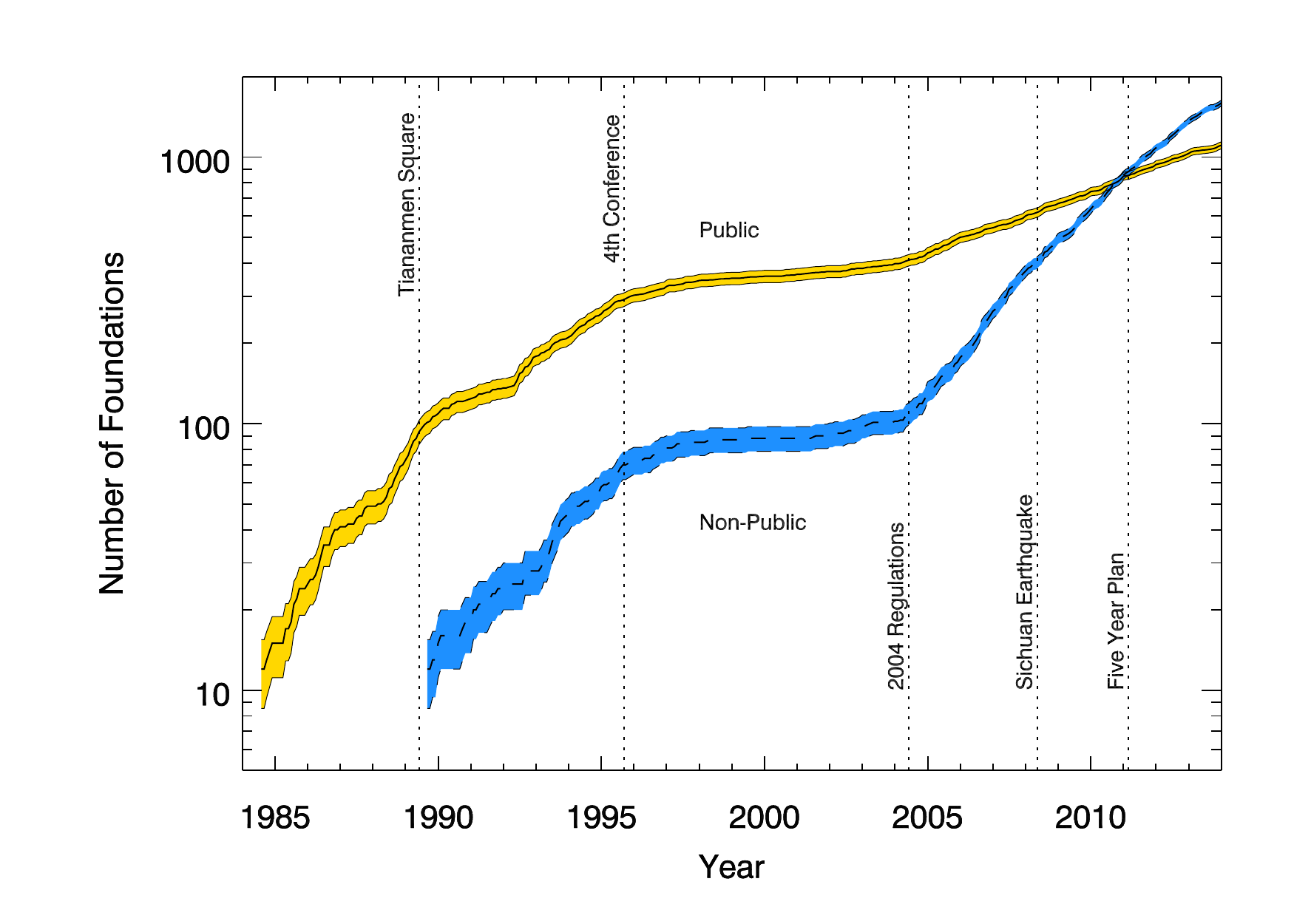} \\
\includegraphics[height=2in]{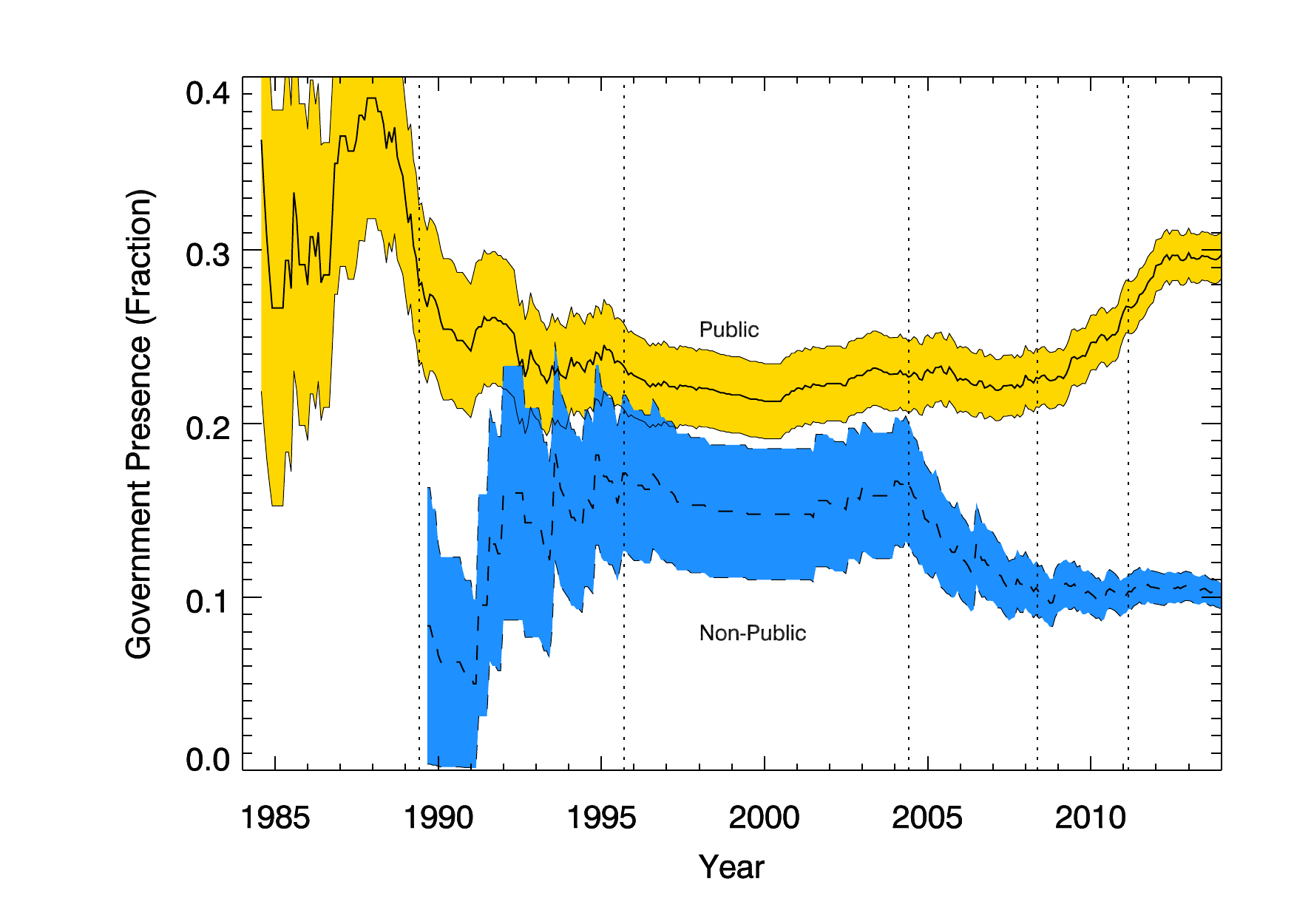} \\
\includegraphics[height=2in]{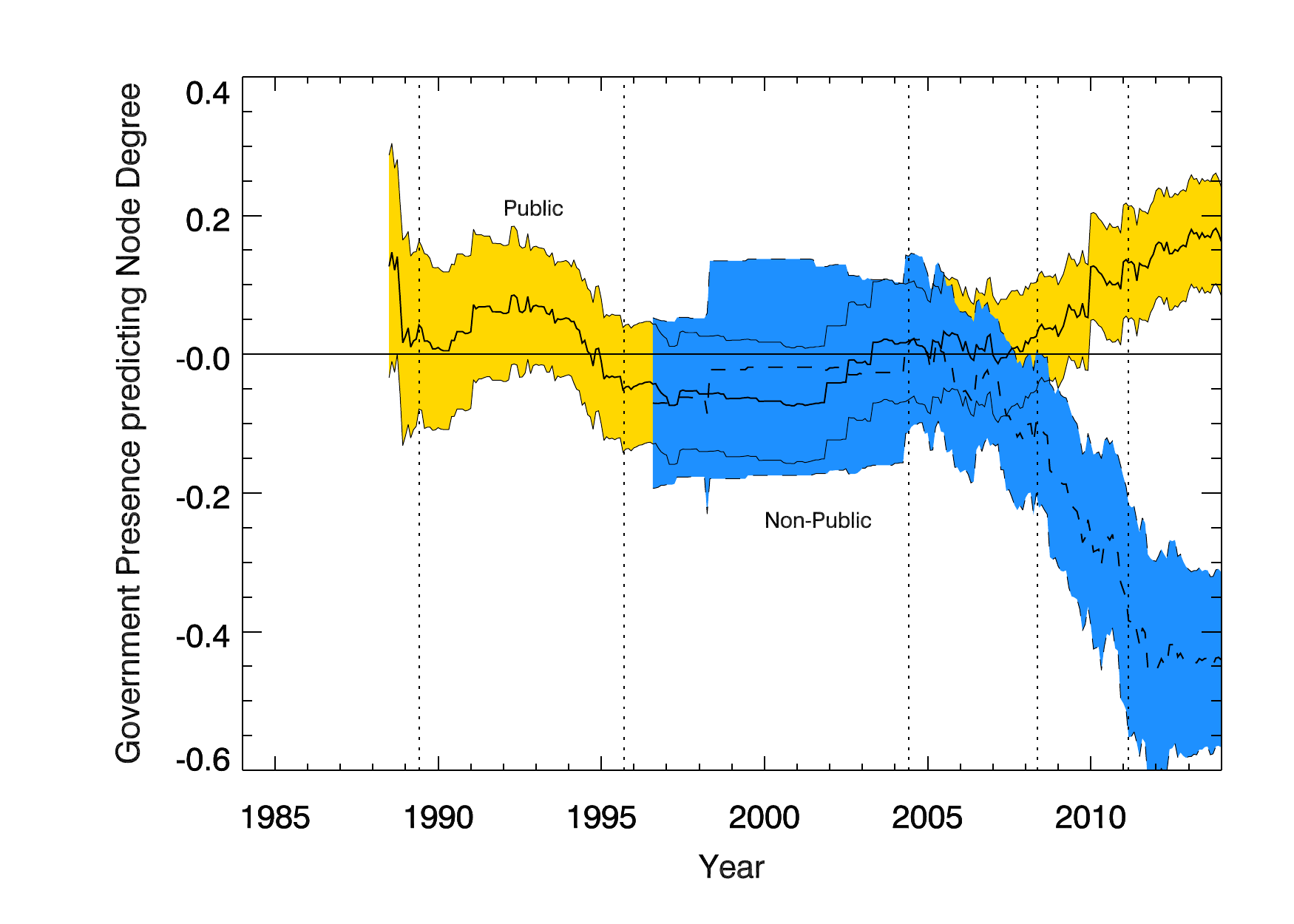}
\end{center}
\caption{Network evolution over time. As the non-public network grew (top panel), it reported lower levels of government presence (middle panel); this contrasts with the growth of the public network which began to report increasing levels of government presence in the same period. In the years after 2008 (see text), the formation of new non-profits leads to the emergence of a detectably negative relationship between node degree and government presence in the non-public network (bottom panel): the highest degree nodes now appear to preferentially exclude government presence. \label{emergence}}
\end{figure}
Fig.~\ref{emergence} shows the three critical metrics over time, under these assumptions: (1) the size of the public and non-public network over time; (2) the changing levels of government presence on boards; and (3) the evolution of the node degree--government presence relationship.

\begin{figure}
\begin{center}
\includegraphics[height=2in]{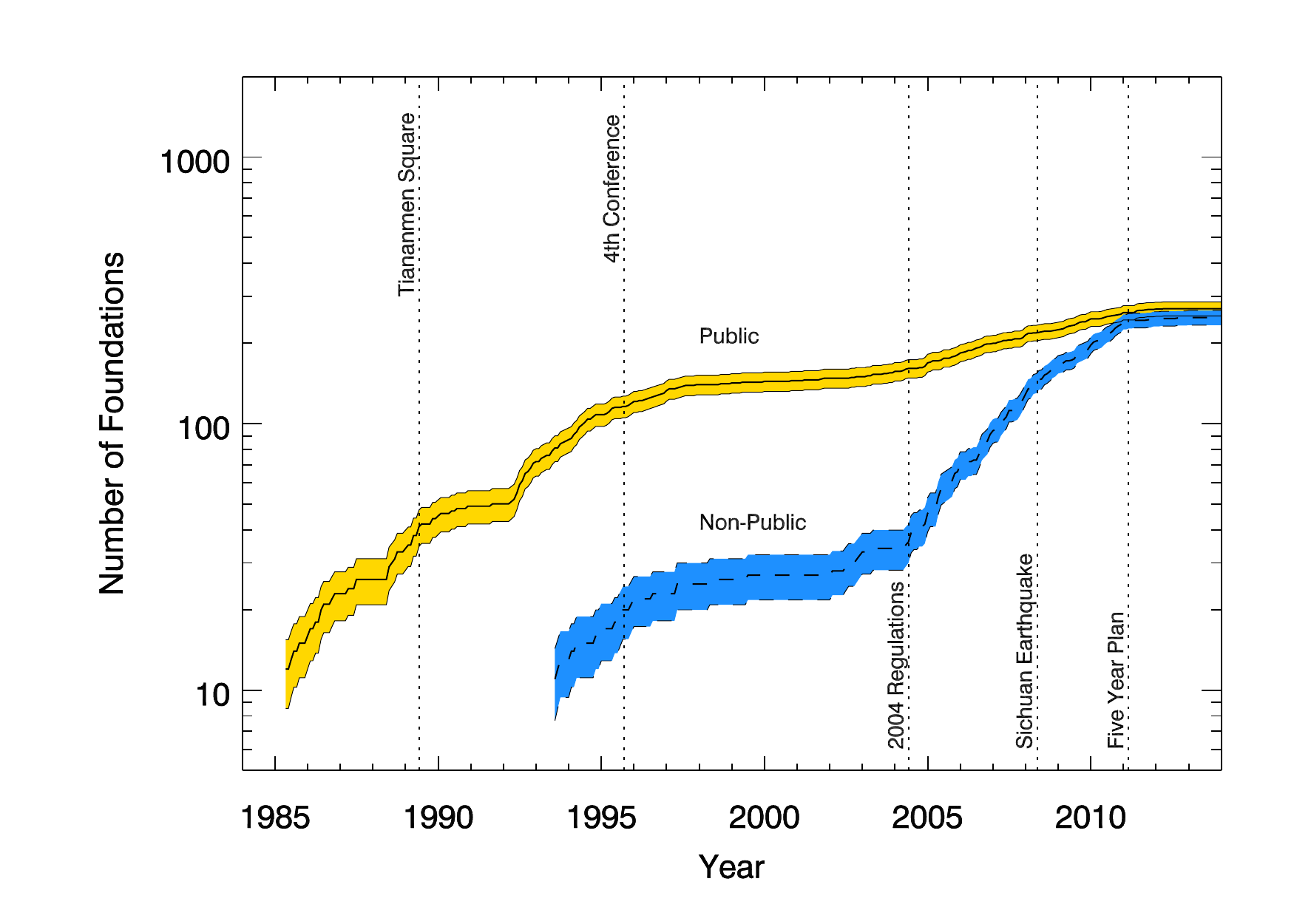} \\
\includegraphics[height=2in]{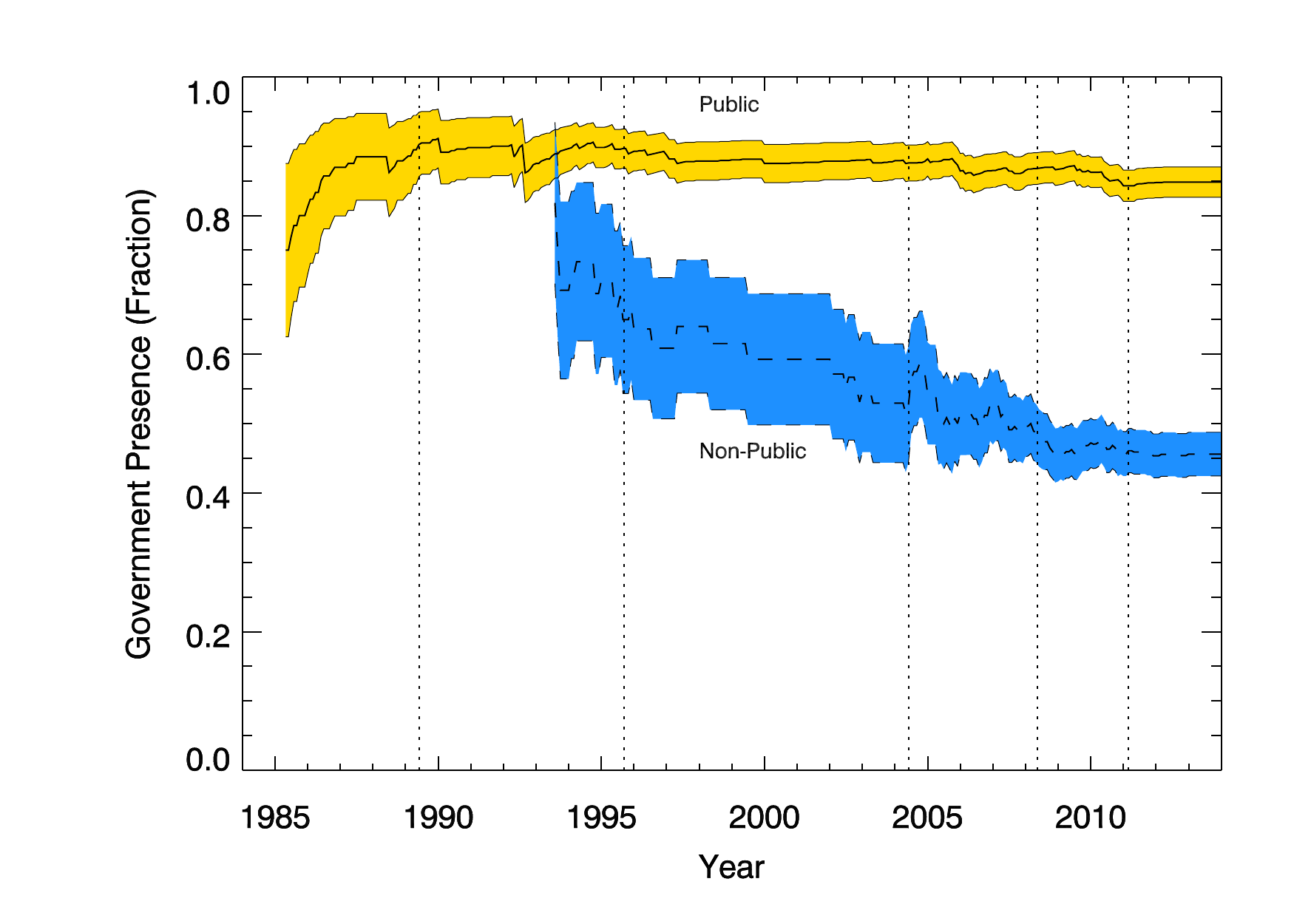} \\
\includegraphics[height=2in]{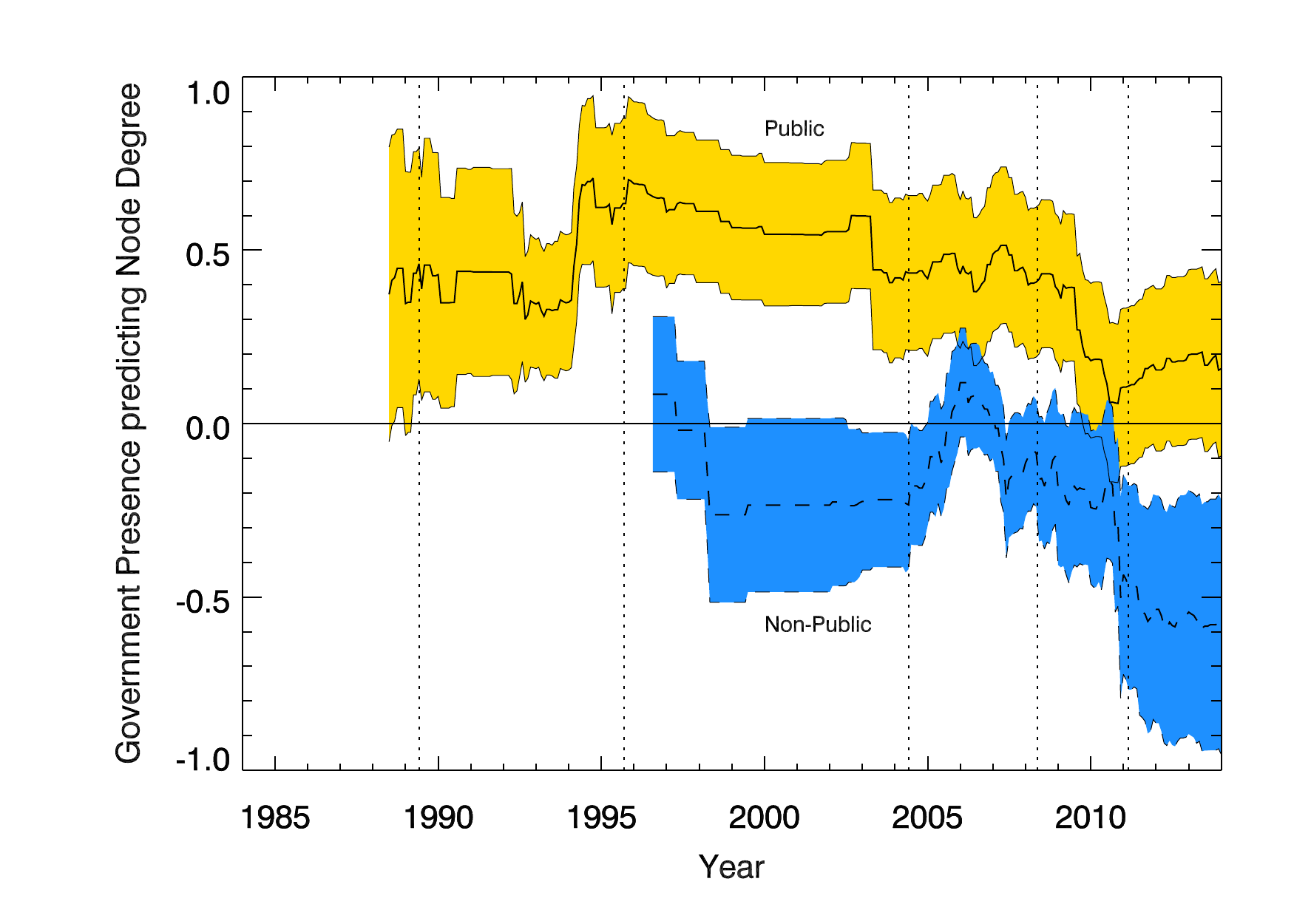}
\end{center}
\caption{Network evolution over time; hand-coded 3A+ subset. Differences in overall levels of supervision are do both to reporting effects, and the necessary inclusion of retired officials in this set. Despite these coding differences, we see the same emergence of elite autonomy signal (bottom panel) as in the full network; here the signal becomes detectably negative after the introduction of the 2011 ``Five Year Plan''. \label{emergence_3a}}
\end{figure}
Overlaid on these panels are five critical dates in the recent history of the non-profit sector: (1) 4 June 1989, the Tiananmen Square protests, and the emergence of civil society in China~\cite{ma2002governance}; (2) 15 September 1995, the Fourth World Conference on Women, associated with the accelerating development of civil society in China~\cite{ma2005non}; (3) 1 June 2004, the day the Regulations on the Management of Foundations, including the above-discussed Article 23, took effect; (4) the Sichuan Earthquake of 12 May 2008, according to Ref.~\cite{shieh_emerging_2011}, a critical date in the expansion of the non-profit sector's horizontal linkages; and (5) the 12th Five-Year plan for 2011--2015, introduced March 2011, which devoted a full chapter to ``social management innovation'' as a key government target. Fig.~\ref{emergence_3a} shows the same plots, but now for the 3A+ hand-coded subset, as discussed above. 

Significant numbers of non-public foundations appear soon after the Tiananmen Square protests, though none of the foundations from that year retain 3A+ status today. The effect of the 2004 Regulations on the growth rate is clear: for at least the next five years, it led to literally exponential growth in the number of both public and non-public foundations. The growth rate in the non-public sector was significantly stronger, so that, by 2011, the non-public sector was larger than the public one. 

The effect of these same regulations on the levels of government supervision is more complex. The 2004 regulations meant that new organizations reported lower levels of government principals in the non-public network. This is consistent with new foundations becoming aware of, and responding to, the restrictions of Article 23. This decline does not appear as strongly in the 3A+ hand-coded subset, however, which suggests one of two explanations: that new foundations are continuing to include government officials, but staying within the letter of the law by including only retired officials, or, that new foundations are continuing to include currently-serving government officials, but under-reporting their presence in official paperwork. The effects of Article 23 are inconsistent: the decline in reported government presence does not occur in the public network; indeed, after 2008, reported government presence starts to rise.

Most interesting are the effects of network growth on the relationship between node degree and government presence. In both the full network, and the 3A+ hand coded subset, we see that organizations responsible for the current negative relationship between degree and government presence joined only late in the network history. Today the non-public network acts to systematically exclude government officials from high degree nodes.  In the full network, the negative relationship becomes strong enough to detect soon after the Sichuan earthquake. In the 3A hand-coded subset, the downward trend towards increasingly negative relationships between degree and government presence becomes detectable only after the 2011 Five Year Plan.

\section{Discussion}

Our quantitative work here confirms the unusual nature of the state-society relationship in contemporary China, one that appears to give significant network powers to independent, non-governmental agents. There are high levels of network clustering and autonomy, and a strong, inverse relationship between network centrality and government supervision. The world of Chinese non-profits is not simple one of command-and-control, where governmental agents dictate by direct presence the actions of the most important players in the network. The nature and strength of these ties may allow for the kinds of decentralized decision-making and policy influence seen in a recent qualitative study of civil society organizations in the country~\cite{teets_power_2017}. 

At the same time, our analysis shows that these foundations are hardly a free-wheeling sector beyond the reach of government power. Our most basic results confirm the persistent and high levels of state presence throughout the non-profit world. The appointment of government officials to high-level positions on foundation boards means that the government continues to hold significant levels of direct influence. In Tocqueville's account of civil society in the 19th Century United States, citizens formed voluntary associations independent of the government itself. Whatever they are doing, the non-profits of 21st century China are far from this 19th Century model, and far, also, from models of post-Soviet democratization in Eastern Europe~\cite{walzer_1992}. Our quantitative results fit with what is widely seen in case studies and fieldwork in China itself: state power may not be complete, but it cannot be ignored.

\subsection{Inconsistent Control}

Our results show significant differences in the level of governmental penetration of the non-profit sector. This is most apparent among the non-public foundations where, at least officially, only one in ten foundations have a current government official as board principal. The government's relationship to its foundations is bimodal, with public foundations showing the highest levels of government involvement; not surprisingly, of this more supervised group, the public foundations engaged in politically-sensitive topics receive the greatest levels of supervision of all.

The unusual nature of these less-supervised, non-public foundations becomes apparent at the network level. A set of network elites are at the center of the non-public network. These elites not only serve as hubs for large numbers of other foundations, but preferentially connect to each other. Examination of the profiles of the highest degree nodes shows that, in a parallel to the United States~\cite{ssq_2002}, the nodes responsible for these interconnections are, primarily, the business elites. 

Detection of this rich club effect provides a new view on the development of horizontal relationships that can enable communication and cooperation---a process captured in fieldwork studies of horizontal connections during the 2008 Sichuan Earthquake~\cite{shieh_emerging_2011}, and a missing piece for understanding the development of state that increasingly devolves government functions to non-government agencies~\cite{saich_2000}. 

That business elites play such a central role in the NGO network fits with recent work that shows an increasing importance of non-corporate venues as corporate networks fragment~\cite{domhoff2009rules, carroll2010making, deGraaff01082012}. Our results go beyond a simple demonstration of elite homophily, however, in showing how elites in an authoritarian regime manage relationships with government officials.

In particular, our network analysis reveals an unexpected negative relationship between node degree and government presence in the non-public network. Not only is government penetration much lower overall in the non-public foundations, it appears to have even less penetration to the foundations at the network's center.  Horizontal relationships between foundations appear to selectively exclude the top-down ``vertical'' control of the state in a form of structural autonomy.

These horizontal relationships are arranged in such a way to create a multipolar structure at the heart of the foundation network. Rather than a set of core-periphery relationships, the network's high levels of clustering suggest that the world of Chinese NGOs is fundamentally multipolar~\cite{heemskerk2016corporate}. Distinct clusters appear to share central places in the network, and the largely equal share of network power given to each cluster means that there is little evidence of a dominant hegemon to which all other clusters uniquely attach. What inequality in network power exists appears, if anything, to be associated with reduced levels of government presence: the two largest clusters in the network have significantly lower levels of government presence than expected.

\subsection{Emergence of Elite Autonomy}

A dynamical analysis of network formation, shows how recent the creation of this partially autonomous sphere has been. Negative relationships between network power and government control appeal soon after the 2008 Earthquake (the downward trend in the 3A subset appears around this time as well, but only becomes statistically significant in 2011). This result lends new support to the qualitative studies of Ref.~\cite{shieh_emerging_2011}, which associated the earthquake to lasting cultural, political, and social changes in the non-profit world that led to new forms of autonomous civil society. Ref.~\cite{zhang2010corporate} supports these results from a different angle, suggesting the increasing importance of private, as opposed to state-controlled, organizations in non-profit relief. In a comparative of study of how government-controlled and private firms responded to the 2008 Earthquake, they found that private firms donated to relief more rapidly, and at greater levels of financial assistance, that state-owned firms. Even many years after the 2008 Earthquake, these effects persist.

One more recent political change comes from the government: the promotion of ``social management innovation'' (\emph{chuangxin shehui guanli}), a core concept of state policy first introduced in the 4th Plenum of the 16th CCP Congress in 2004~\cite{pieke_communist_2012, schlaeger_official_2014}; the 12th Five-Year plan for 2011--2015 elaborated the policy as a governance target, proposing ``party leadership, government responsibility, societal cooperation, public participation'' as key principles~\cite{translation}, the use of civil society to provide social services, and a correspondingly less rigid approach to controlling it~\cite{shan_social_2013}. Our results confirm the impact of these further reforms: in the post-plan era, the negative relationship between government presence and node degree has persisted. 

Crucially, our data does not include the so-called ``grassroots'', or illegal NGOs that, while tolerated by the state, operate without legal sanction~\cite{illegal_ngo}. These NGOs sometimes include Communist party members on staff, are largely tolerated by the state and often seen as socially legitimate organizations, and often provide essential services in cases where the state can not. Despite their technical illegality, government officials sometimes even lend explicit support to their mission. This qualitative work further suggests that members of illegal NGOs are  largely uninterested in organizing independently in ways that might appear to threaten the state. Taken at face value, these suggest that were data on the staff or board members of illegal NGOs to be incorporated into our networks, we would find them in a core-periphery relationships to the officially-sanctioned actors we study here. 

\subsection{Civil Society and the Meaning of Network Connectivity}

Network connectivity is associated with the ability to form common knowledge~\cite{chwe1999structure}, and both simulations and theory suggest that nodes of high degree play a crucial role when common knowledge is required for joint action~\cite{liddell2015common}. The negative relationship between node degree and government presence observed in our data then amounts to a form of structural autonomy. Taken at face value, these results seem to suggest that foundations within this non-public subset may be part of the emergence of a new civil society, whose interlocks occur increasingly independently of the state, and whose resultant capacity for independent coordination might even be seen as a threat to authoritarian control. 

The identification of these links with a Tocquevillian civil society is complicated, however, by the fact that when the most connected of the non-public foundations are not government in origin, they appear, instead, to be drawn from the business elite. The extensive ties between the state and business then suggest that this civil society is something less than might be expected; ever since the reforms of the 1980s, scholars have suggested that the business elite act as an agent of the government itself~\cite{pearson_chinas_1997}. 

Indeed, this provides a clear alternate explanation: if the highest degree nodes are sufficiently aligned with the government to begin with, they can be allowed to operate without direct supervision---precisely as we see in our data. Whether or not the elite are truly independent of the government, their special position in society, and their ability to influence the government financially~\cite{ma_tocquevillian_2006} suggests that this civil society, such as it is, is far from the pluralist Tocquevillian world of the ordinary citizen. Understood in this context, and with the caveat that our work here is able to cover only the official NGOs, and not the illegal sector, these results are perhaps more consistent with the theory of ``consultative authoritarianism'', argued for by Ref.~\cite{teets_2013}, in which the government tolerates increasing levels of autonomy among non-governmental organizations while developing new strategies of indirect control. Networks of state power may overlap with ones defined by economic exchange, but need not be coextensive~\cite{mann1986sources}. 

Our results show that the government has come to tolerate communities of interlocking associations that operate with lower levels of direct government presence. Today, the total number of these organizations is small. The new Charity Law, effective from 1 September 2016, allows a far greater number of associations---potentially in the hundreds of thousands---to raise funds from the general public. We expect the rapid expansion in both the size and scale of the Chinese non-profit sector will radically increase its impact on the state-society relationship. 

\subsection{Networked Civil Society and Multipolar Structure}

Finally, our cluster analysis suggests that the social actors in civil society are organized into larger clusters by horizontal relationships, and arranged in such a way to creature a relatively equal relationship between the most important clusters in the network: a multipolar world~\cite{heemskerk2016corporate} of Chinese NGOs. Distinct clusters appear to share central places in the network, and the equal shares of network power given to each cluster is very different from a system with a dominant hegemon to which all other clusters uniquely attach. 

The clusters are not perfectly equal in network power. The inequality in network power that does exist appears to be associated with reduced levels of government presence: the two largest clusters in the network have significantly lower levels of government presence than expected. This result, concerning the most powerful clusters of individual nodes, parallels the results on elite autonomy among the nodes with greatest network power on the individual level.

By allowing these more egalitarian, more horizontal structures to emerge---and, further, by allowing the most important of these structures to operate with increased independence from central monitoring---the state may have made it possible for organizations to make more efficient use of local and distributed knowledge~\cite{hayek_1945}. The patterns in which these new structures arrange themselves call to mind classic theories of decentralized decision-making in liberal societies that draw attention to autonomous, spontaneous, and ``polycentric'' orders~\cite{polanyi_logic_1951,hayek_constitution_2011}. At the same time, persistently high levels of state involvement in the sector frustrate a simple analogy to descriptions of liberal states. Our account of Chinese civil society here, in terms of the simultaneous presence of both horizontal (group-to-group), and vertical (group-to-state) linkages, and the structural tensions between the two, finds parallels in recent work on collective action in rural China~\cite{lu2017organizational}, which identifies a new class of ``semi-integrated'' organizations that can mobilize both horizontal and vertical linkages at once.

Remarkably, the Chinese government does not appear to be combating the emergence of a potential civil society counter-power either through direct monitoring, or through disruption of network ties. There is plenty of evidence that individual foundations in the official Chinese non-profit sector today are relatively tame, with a strong tendency to align their goals with that of the state~\cite{teets_2013}. Yet the underlying network structures that these organizations have implicitly created have the potential to enable more independent action and decision-making than one would expect; they are certainly very far away from the highly centralized social structure and planned economy that characterized mainland China during Mao's rule. Our work suggests that the government will either increasingly employ less-visible strategies of control, or, conversely, come to accept the delivery of social and economic needs through a rapidly-growing, complex, and increasingly autonomous sphere.

\clearpage 
\section*{Acknowledgments}
\noindent
J.M.\ thanks Dr. Zhaonan Zhu and Bin Chen for assistance with coding; Qun Wang and the RICF data quality team for data collection and data quality control. S.D.\ thanks the Santa Fe Institute and the Alan Turing Institute for their hospitality while this work was completed. We thank Dr.\ Peter Frumkin and the 2016 Penn Summer Doctoral Fellows Program fellows, Dr. Richard Steinberg, Dr. Lehn Benjamin, Dr. Bin Chen, Dr. Xinsong Wang, Xunyu Xiang, Xiaoyun Wang, Bradi Heaberlin, and Torrin Liddell for their valuable comments. The RICF project is supported in part by the Dunhe Foundation.

\clearpage 
\section*{References}

\end{document}